\documentclass{aa}  
\usepackage{graphicx}
\usepackage{txfonts}
\newcommand{\va}{v_{\mathrm{A}}}
\newcommand{\cs}{c_{\mathrm{s}}}

\newcommand{\nct}{\tilde{c}_{\mathrm{T}}}

\newcommand{\vap}{v_{\mathrm{Ap}}}
\newcommand{\csp}{c_{\mathrm{sp}}}

\newcommand{\tildectp}{\tilde{c}_{\mathrm{Tp}}}

\newcommand{\lambdap}{\Lambda_{\mathrm{p}}}
\newcommand{\lambdac}{\Lambda_{\mathrm{c}}}

\newcommand{\vac}{v_{\mathrm{Ac}}}
\newcommand{\csco}{c_{\mathrm{sc}}}
\newcommand{\ctc}{c_{\mathrm{Tc}}}

\newcommand{\kzp}{k_{z \mathrm{p}}}
\newcommand{\kzc}{k_{z \mathrm{c}}}

\begin{document}

	\title{The effect of the solar corona on the attenuation of small-amplitude prominence oscillations}

	\subtitle{I. Longitudinal magnetic field}

   \author{R. Soler \and R. Oliver \and J. L. Ballester}

   \offprints{R. Soler}

   \institute{Departament de F\'isica, Universitat de les Illes Balears,
              E-07122, Palma de Mallorca, Spain\\
              \email{[roberto.soler;ramon.oliver;dfsjlb0]@uib.es}
             }

   \date{Received xxxx; accepted xxxx}

  \abstract
   {One of the typical features shown by observations of solar prominence 
oscillations is that they are damped in time and that the values of the 
damping times are usually between one and three times the corresponding 
oscillatory period. However, the mechanism responsible for the attenuation is 
still not well-known.}
   {Thermal conduction, optically thin or thick radiation and heating are 
taken into account in the energy equation, and their role on the attenuation 
of prominence oscillations is evaluated.}
   {The dispersion relation for linear non-adiabatic magnetoacoustic waves is derived considering an equilibrium made of a prominence plasma 
slab embedded in an unbounded corona. The magnetic field is orientated along 
the direction parallel to the slab axis and has the same strength in all 
regions. By solving the dispersion relation for a fixed wavenumber, a complex 
oscillatory frequency is obtained, and the period and the damping time are 
computed.}
   {The effect of conduction and radiation losses is different for each 
magnetoacoustic mode and depends on the wavenumber. In the observed range of 
wavelengths the internal slow mode is attenuated by radiation from the 
prominence plasma, the fast mode by the combination of prominence radiation 
and coronal conduction and the external slow mode by coronal conduction. The 
consideration of the external corona is of paramount importance in the case 
of the fast and external slow modes, whereas it does not affect the internal 
slow modes at all. When a thinner slab representing a filament thread is considered the fast mode is less attenuatted whereas both internal and external slow modes are not affected.}
   {Non-adiabatic effects are efficient damping mechanisms for magnetoacoustic 
modes, and the values of the obtained damping times are compatible with those 
observed.}


	\titlerunning{The effect of the solar corona on the attenuation of prominence oscillations I}

	       
   \maketitle
%


\section{Introduction}

Prominences are dense coronal structures which appear as thin, dark filaments on the solar disc when observed in H$\alpha$. On the contrary, they show up as bright objects above the solar limb. The coronal magnetic field is responsible for the support of prominences against
gravity, and it also plays a fundamental role in the thermal confinement of the
cool prominence plasma embedded in the much hotter coronal environment. Nevertheless, the structure, orientation and strength of the magnetic field in prominences and the surrounding corona is still enigmatic and not well-known. High resolution observations reveal that prominences are composed by numerous very thin, thread-like structures, called fibrils, piled up to form the body of the prominence (Lin et al. 2003; Lin et al. 2005, Lin et al. 2007) and measures also indicate that magnetic field lines are orientated along these thin threads.

The observational evidence of small-amplitude oscillations in quiescent solar
prominences goes back to 40 years ago (Harvey 1969). The amplitude of these oscillations typically goes from less than 0.1~km~s$^{-1}$ to 2--3~km~s$^{-1}$, and have been
historically classified, according to their periods, in short- ($P <
10$~min), intermediate- ($10$~min~$< P <$~$40$~min) and long-period oscillations
($P > 40$~min), although very short-periods of less than 1~min (Balthasar et al.
1993) and extreme ultra-long-periods of more than 8~hours (Foullon et al. 2004)
have been reported. Nevertheless, the value of the period seems not to be related with the
nature or the source of the trigger and probably is linked to the prominence eigenmode that is excited. There are also a few determinations of the wavelength and phase speed of standing oscillations and propagating waves in large regions of prominences (Molowny-Horas et al. 1997; Terradas et al. 2002) and in single filament threads (Lin et al. 2007). On the other hand, several observations (Molowny-Horas et al. 1999; Terradas et al. 2002) have informed about the evidence of the attenuation of the oscillations in Doppler velocity time series, which is a common feature observed in large areas. By fitting a sinusoidal function multiplied by a factor $\exp (-t/\tau_{\rm D})$ to the Doppler series, these authors have obtained values of the damping time, $\tau_{\rm D}$, which are usually between 1 and 3 times the corresponding oscillatory period. The reader is referred to some recent reviews for more information about the observational background  (Oliver \& Ballester 2002, Wiehr 2004, Engvold 2004, Ballester 2006). 

 From the theoretical point of view, small-amplitude prominence oscillations can be interpreted in terms of linear magnetohydrodynamic (MHD) waves. Although, there is a wide bibliography of works that investigate the ideal MHD wave modes supported by prominence models (see Oliver \& Ballester 2002 for an extensive review of theoretical studies), the investigation of the wave damping has been broached in few papers. By removing the ideal assumption and including dissipative terms in the basic MHD equations, several works have studied the attenuation of prominence oscillations considering radiative losses based on the Newtonian law of cooling with a constant relaxation time (Terradas et al 2001), or performing a more complete treatment of non-adiabatic effects, assuming optically thin radiation, heating and thermal conduction (Carbonell et al. 2004; Terradas et al. 2005). The main conclusion of these previous studies is that only the slow wave is attenuated by thermal effects, radiation being the dominant damping mechanism in the range of typically observed wavelengths in prominences, but the fast wave remains practically undamped. On the other hand, Forteza et al. (2007) proposed ion-neutral collisions as a damping mechanism on the basis that prominences are partially ionised plasmas, but they found that this mechanism is only efficient in attenuating the fast mode in quasi-neutral plasmas, the slow mode being almost unaffected.    

 In the light of these referred studies, it is likely that non-adiabatic effects are the best candidates for the damping of small-amplitude oscillations, at least for slow modes. However, previous results do not asses the influence of the corona. The main aim of the present work is to perform a step forward in the investigation of the effect of non-adiabatic mechanisms (radiation losses, thermal conduction and heating) on the time damping of prominence oscillations. We consider a slab model with a longitudinal magnetic field and take into account the external coronal medium. So, we explore for the first time the joint effect of prominence and coronal mechanisms on the attenuation of oscillations. The magnetoacoustic normal modes of this equilibrium have been previously investigated by Edwin \& Roberts (1982) and Joarder \& Roberts (1992) in the adiabatic case. Later, a revision of these works has been done in Soler et al. (2007), hereafter Paper~I, and the normal modes have been studied and reclassified according to their magnetoacoustic properties.

This paper is organised as follows. The description of the equilibrium model and the linear non-adiabatic wave equations are given in Sect.~\ref{sec:equations}, whereas the dispersion relation for the magnetoacoustic modes is derived in Sect.~\ref{sec:dispersion}. Then, the results are plotted and investigated in Sect.~\ref{subsec:results}. Finally, Sect.~\ref{sec:conclusions} contains the conclusions of this work.


\section{Equilibrium and basic equations \label{sec:equations}} 

\begin{figure}[!htb]
\centering
\includegraphics[width=0.95\columnwidth]{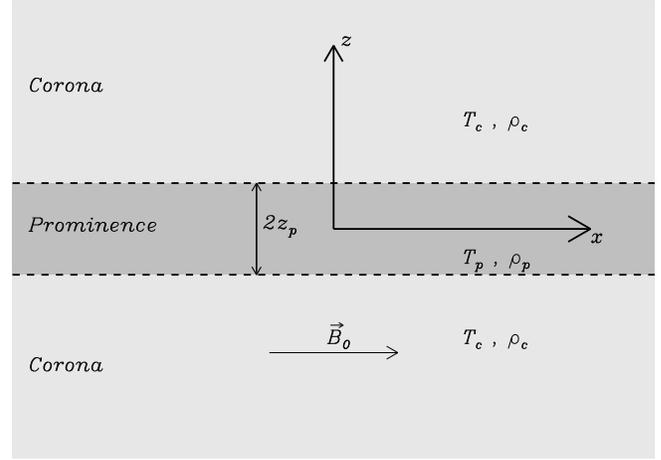}
\caption{Sketch of the equilibrium. \label{fig:equilibrium}}
\end{figure}

Our equilibrium configuration (Fig. \ref{fig:equilibrium}) is made of a homogeneous plasma layer with prominence conditions (density $\rho_{\rm p}$ and temperature $T_{\rm p}$) embedded in an unbounded corona (density $\rho_{\rm c}$ and temperature $T_{\rm c}$). The coronal density is computed by fixing the coronal temperature and imposing pressure continuity across the interfaces. The magnetic field is $\vec{B}_0=B_0 \hat{e}_x$, with $B_0$ a constant everywhere. Both media are unlimited in the $x$- and $y$-directions. The half-width of the prominence slab is $z_{\rm p}$. 

The basic magnetohydrodynamic equations for the discussion of non-adiabatic processes are:

\begin{equation}
\frac{D \rho}{D t} + \rho \nabla \cdot \vec{v} = 0, \label{1}
\end{equation}
\begin{equation}
\rho \frac{D \vec{v}}{D t} = - \nabla p + \frac{1}{\mu}
(\nabla \times \vec{B}) \times \vec{B}, \label{2}
\end{equation}
\begin{equation}
\frac{D p}{D t} - \frac{\gamma p}{\rho} \frac{D \rho}{D t} + (\gamma - 1)[\rho L(\rho,T)-\nabla \cdot (\vec{\kappa} \cdot \nabla T)] = 0, \label{3}
\end{equation}
\begin{equation}
\frac{\partial \vec{B}}{\partial t} = \nabla \times (\vec{v} \times \vec{B}), \label{4}
\end{equation}
\begin{equation}
\nabla \cdot \vec{B} = 0, \label{5}
\end{equation}
\begin{equation}
p = \frac{\rho R T}{\tilde{\mu}}, \label{6}
\end{equation}
where $\frac{D}{D t} = \frac{\partial}{\partial t} + \vec{v} \cdot \nabla$ is the material derivative for time variations following the motion and all quantities have their usual meaning.  Equation~(\ref{3}) is the energy equation, which in the present form takes into account non-adiabatic effects (radiation losses, thermal conduction and heating) and whose terms are explained in detail in Carbonell et al. (2004) and Terradas et al. (2005). Following these works, only thermal conduction parallel to the magnetic field is assumed and we use the typical value for the parallel conductivity in prominence and coronal applications, $\kappa_\parallel = 10^{-11} T^{5/2}\, \mathrm{W}\, \mathrm{m}^{-1}\, \mathrm{K}^{-1}$. Radiative losses and heating are evaluated together through the heat-loss function, $L(\rho,T)= \chi^* \rho T^\alpha - h \rho^a T^b$, where radiation is parametrised with $\chi^*$ and $\alpha$ (see Table~\ref{tab:parameters}) and the heating scenario is given by exponents $a$ and $b$. The heating mechanisms taken into account in this work are (Rosner et al. \cite{rosner}; Dahlburg \& Mariska \cite{dahlburg}):
\begin{itemize}
\item constant heating per unit volume ($a=b=0$);
\item constant heating per unit mass ($a=1$, $b=0$); 
\item heating by coronal current dissipation ($a=b=1$);
\item heating by Alfv\'en mode/mode conversion ($a=b=7/6$);
\item heating by Alfv\'en mode/anomalous conduction damping ($a=1/2$, $b=-1/2$).
\end{itemize}

\begin{table}
\caption{Parameter values of the radiative loss function corresponding to the considered regimes. The three prominence regimes represent different plasma optical thicknesses. Prominence~(1) regime corresponds to an optically thin plasma whereas Prominence~(2) and Prominence~(3) regimes represent greater optical thicknesses. All quantities are expressed in MKS units.}             
\label{tab:parameters}      
\centering                          
\begin{tabular}{l c c l}        
\hline\hline                 
Regime & $\chi^*$ & $\alpha$ & Reference \\    
\hline                        
   Prominence~(1) & $1.76 \times 10^{-13}$ & $7.4$  & Hildner (1974) \\      
   Prominence~(2) & $1.76 \times 10^{-53}$ &  $17.4$   & Milne et al. (1979) \\
   Prominence~(3) & $7.01 \times 10^{-104}$ &  $30$   & Rosner et al. (1978) \\
   Corona  & $1.97 \times 10^{24}$ & $-1$ & Hildner (1974) \\ 
\hline                                   
\end{tabular}
\end{table}

Following the same process as in Carbonell et al. (2004), we consider small perturbations from the equilibrium state, linearise the basic Eqs. (\ref{1})--(\ref{6}) and obtain their Eqs. (9)--(14). Since our model is unlimited in the $x$- and $y$-directions, we assume all perturbations are in the form $f_1(z)\exp i (\omega t + k_x x + k_y y)$, and considering only motions and propagation in the $xz$-plane ($v_y=0$, $k_y=0$), which excludes Alfv\'en waves, the linearised equations become 
\begin{equation}
i \omega \rho_1 + \rho_0 \left( i k_x v_x + \frac{{\rm d} v_z}{{\rm d} z} \right) = 0, \label{15}
\end{equation}
\begin{equation}
i \omega \rho_0 v_x = - i k_x p_1, \label{16}
\end{equation}
\begin{equation}
i \omega \rho_0 v_z = -\frac{{\rm d} p_1}{{\rm d} z} + \frac{B_0}{\mu} \left( i k_x B_{1z} - \frac{{\rm d} B_{1x}}{{\rm d} z} \right), \label{17}
\end{equation}
\begin{equation}
i \omega \left( p_1 - \cs^2 \rho_1 \right) = -(\gamma-1) \left( k_x^2 \kappa_\parallel T_1 +  \frac{p_0}{\rho_0}\omega_\rho \rho_1 +  \frac{p_0}{T_0} \omega_T T_1 \right), \label{18}
\end{equation}
\begin{equation}
i \omega B_{1x} = -B_0 \frac{{\rm d} v_z}{{\rm d} z}, \label{19}
\end{equation}
\begin{equation}
i \omega B_{1z} = B_0 i k_x v_z, \label{20}
\end{equation}
where $\cs^2 = \frac{\gamma p_0}{\rho_0}$ is the adiabatic sound speed squared and 
\begin{displaymath}
\omega_\rho \equiv \frac{\rho_0}{p_0}\left( L + \rho_0 L_\rho \right), \qquad \omega_T \equiv \frac{\rho_0}{p_0}T_0 L_T,
\end{displaymath}
$L_\rho$, $L_T$ being the partial derivatives of the heat-loss function with respect to density and temperature, respectively,
\begin{displaymath}
L_\rho \equiv \left( \frac{\partial L}{\partial \rho} \right)_T, \qquad 
L_T \equiv \left( \frac{\partial L}{\partial T} \right)_\rho.
\end{displaymath}

Now, it is possible to eliminate all perturbations in favour of $v_z$ to obtain a single differential equation
\begin{equation}
\frac{{\rm d}^2 v_z}{{\rm d} z^2} + k_z^2 v_z = 0, \label{21}
\end{equation}
in which
\begin{equation}
k_z^2 = \frac{\left( \omega^2 - k_x^2 \va^2 \right) \left( \omega^2 - k_x^2 \Lambda^2 \right)}
{\left( \va^2 + \Lambda^2 \right) \left( \omega^2 - k_x^2 \nct^2 \right)}, \label{22}
\end{equation}
where $\va^2 = \frac{B_0^2}{\mu \rho_0}$ is the Alfv\'en speed squared. $\Lambda^2$ and $\nct^2$ are the modified sound and cusp (or tube) speed squared, respectively,
\begin{equation}
\Lambda^2 \equiv \frac{\cs^2}{\gamma} \left[ \frac{\left( \gamma-1 \right) \left( \frac{T_0}{p_0} \kappa_\parallel k_x^2 + \omega_T - \omega_\rho \right) + i \gamma \omega}
{\left( \gamma -1 \right) \left( \frac{T_0}{p_0} \kappa_\parallel k_x^2 + \omega_T \right) + i \omega} \right], \label{23}
\end{equation}
\begin{equation}
\nct^2 \equiv \frac{\va^2 \Lambda^2}{\va^2 + \Lambda^2}. \label{24}
\end{equation}

Expressions for the perturbations in terms of $v_z$ are given in App.~A. In all the following formulae, subscripts p or c denote quantities computed using prominence or coronal values, respectively.

\section{Dispersion relation \label{sec:dispersion}}

We impose some restrictions on the solutions of Eq. (\ref{21}) in order to obtain the dispersion relation for the linear non-adiabatic magnetoacoustic waves. We restrict this analysis to body waves which are evanescent in the corona, since we are looking for solutions which are essentially confined to the slab.  For such solutions, $v_z(z)$ is of the form
\begin{equation}
  v_z(z) = \left\{
  \begin{array}{lcl}
  A_1 \exp \left[ \kzc \left( z + z_{\rm p} \right) \right], & \mathrm{if} & z \leq -z_{\rm p}, \\
  A_2 \cos \left( \kzp z \right) + A_3 \sin \left( \kzp z \right), & \mathrm{if} & -z_{\rm p} \leq z \leq z_{\rm p}, \\
  A_4 \exp \left[- \kzc \left( z - z_{\rm p} \right) \right], & \mathrm{if} & z \geq z_{\rm p}. 
  \end{array} \right. \label{25}
\end{equation}
with $ \Re (\kzp ) > 0 $ and  $ \Re (\kzc ) > 0 $.

Imposing continuity of $v_z$ and the total (gas plus magnetic) pressure perturbation across the interfaces, we find four algebraic relations between the constants $A_1$, $A_2$, $A_3$ and $A_4$. The non-trivial solution of this system gives us the dispersion relation
\begin{equation}
\frac{\rho_{\rm c}}{ \rho_{\rm p}} \left( k_{x}^{2} \vac^2 - \omega^{2}\right) \kzp 
\left\{ \begin{array}{l} \cot \\ \tan \end{array} \right\}  
\left( \kzp z_{\rm p} \right) \pm \left( k_{x}^{2} \vap^{2} - \omega^{2}\right) \kzc =0, \label{26}
\end{equation}
where $\cot$/$\tan$ terms and $\pm$ signs are related with the symmetry of the perturbations. The $\cot$ term and the $+$ sign correspond to kink modes ($A_3=0$), whereas the $\tan$ term and the $-$ sign correspond to sausage modes ($A_2=0$).

The dispersion relation for the magnetoacoustic waves presented in Eq. (\ref{26}) is equivalent to the relation investigated in Edwin \& Roberts (1982) and Joarder \& Roberts (1992), and revised Paper~I, in the case of adiabatic perturbations, since all non-adiabatic terms are now enclosed in $\kzp$ and $\kzc$ through Eq.~(\ref{22}).


\section{Results}

\label{subsec:results}

Now, we assume Prominence~(1) conditions inside the slab (i.e. an optically thin prominence) and a heating mechanism given by \mbox{$a=b=0$}. Unless otherwise stated,  the following equilibrium parameters are considered in all computations: $T_{\rm p}=8000$~K, \mbox{$\rho_{\rm p} = 5 \times 10^{-11}$~kg~m$^{-3}$}, $T_{\rm c} = 10^6$~K, $\rho_{\rm c} = 2.5 \times 10^{-13}$~kg~m$^{-3}$, $B_0=5$~G and $z_{\rm p} = 3000$~km. The solution of the dispersion relation (Eq.~[\ref{26}]) for a fixed real $k_x$ gives us a complex frequency $\omega = \omega_{\rm R} + i \omega_{\rm I}$. We then compute the oscillatory period, the damping time and the ratio of the damping time to the period because this is an important quantity from the observational point of view, 
\begin{displaymath}
P = \frac{2 \pi}{\omega_{\rm R}}, \qquad \tau_{\rm D} = \frac{1}{\omega_{\rm I}}, \qquad \frac{\tau_{\rm D}}{P} = \frac{1}{2\pi} \frac{\omega_{\rm R}}{\omega_{\rm I}}
\end{displaymath}

\begin{figure}[!htb]
\centering
\includegraphics[width=\columnwidth]{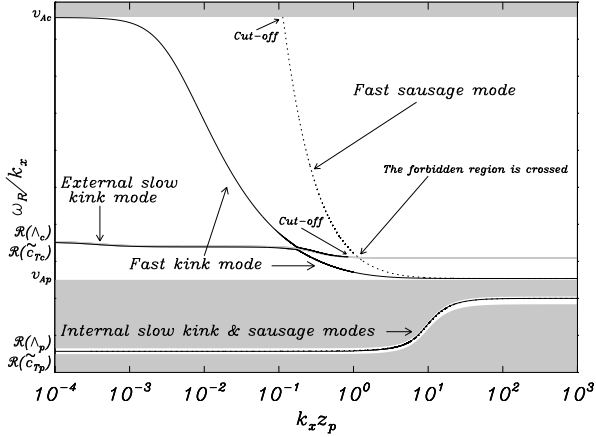}
\caption{Phase speed versus the dimensionless wavenumber for the three fundamental oscillatory modes. Solid lines denote kink modes whereas dotted lines correspond to sausage modes. The shaded zones are projections of the forbidden (or leaky) regions on the plane of this diagram. Note that the vertical axis is not drawn to scale. \label{fig:dispersion}}
\end{figure}

 Since we are interested in studying the behaviour of the most relevant solutions of the dispersion relation, we only compute the results for the fundamental modes, which are labelled, according to the classification of Paper~I, as internal and external slow modes and fast modes. The band structure described in Paper~I is slightly modified when non-adiabatic terms are considered (see Fig.~\ref{fig:dispersion}). The phase speed of the internal slow modes is now enclosed in the band $\Re ( \tildectp ) < \omega_{\rm R}/k_x < \Re ( \lambdap )$. The adiabatic fast modes exist in two separated bands in the phase speed diagram due to the presence of a forbidden region ($\ctc < \omega_{\rm R}/k_x < \csco$), but now the forbidden band is avoided and a continuous fast mode is found with $\vap < \omega_{\rm R}/k_x < \vac$. Finally, and like in the adiabatic case, among the external slow modes only the fundamental kink one exists as a non-leaky solution in a restricted wavenumber range and couples with the fundamental fast kink mode. Its phase speed is $\omega_{\rm R}/k_x \approx \Re ( \lambdac )$. Therefore, we see that in the non-adiabatic case $\Lambda$ plays the role of $\cs$ in the adiabatic case.

\begin{figure*}[!htp]
\centering
\includegraphics[width=2\columnwidth]{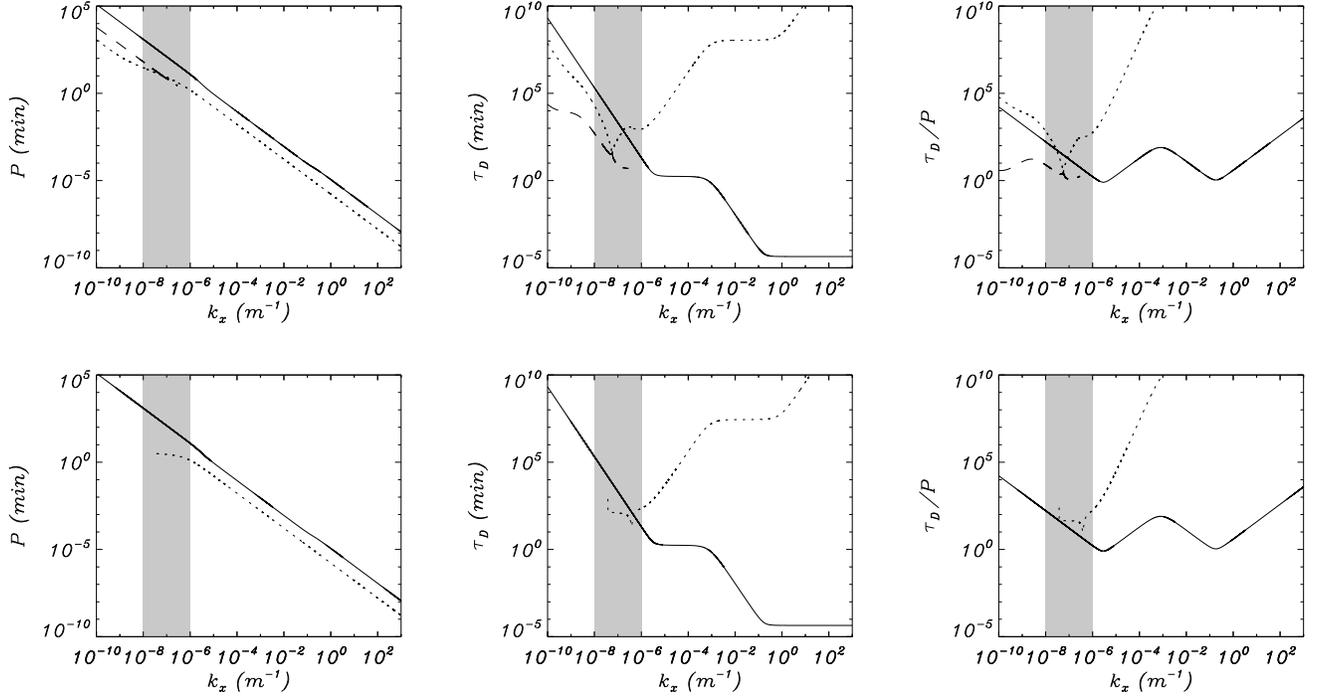}
\caption{Period (left), damping time (centre) and ratio of the damping time to the period (right) versus the longitudinal wavenumber for the fundamental oscillatory modes. Upper panels: internal slow kink (solid line), fast kink (dotted line) and external slow kink (dashed line). Lower panels: internal slow sausage (solid line) and fast sausage (dotted line). Shaded zones correspond to those wavelengths typically observed. Note the cut-offs of the external slow kink mode and the fast sausage mode. Prominence~(1) radiation conditions have been taken for the prominence plasma and the heating scenario is given by $a=b=0$. \label{fig:nonadgen}}
\end{figure*}

In Fig.~\ref{fig:nonadgen} $P$, $\tau_{\rm D}$ and $\tau_{\rm D}/P$ are represented for the fundamental modes and for a range of the longitudinal wavenumber between $10^{-10}\, \mathrm{m}^{-1}$ and $10^{3}\, \mathrm{m}^{-1}$. The shaded zones correspond to wavelengths between $5\times 10^{3}\, \mathrm{km}$ and $10^{5}\, \mathrm{km}$, the typically observed values. It turns out that the values of the period are very similar to those obtained in the adiabatic case (Joarder \& Roberts \cite{joarder}; Paper~I). The damping time presents a strong dependence with the wavenumber and its behaviour is very different from one mode to another. This fact suggests that the non-adiabatic mechanisms can affect each mode in a different way (Carbonell et al. 2004). This is studied in detail in Sect~\ref{subsec:mechanism}. Observations show that prominence oscillations are typically attenuated in a few periods (Terradas et al. 2002), so a damping time of the order of the period is expected. In our results, the fundamental modes present values of $\tau_{\rm D}/P$ in the range 1 to 10 in the observed wavelength region, which is in agreement with observations.


\subsection{Regions of dominance of the damping mechanisms}
\label{subsec:mechanism}

The importance of the different non-adiabatic terms included in the energy equation (Eq.~[\ref{3}]) depends on the wavenumber. In order to know which is the range of dominance of each mechanism, we compare the damping time obtained when considering all non-adiabatic terms (displayed in the middle column of Fig.~\ref{fig:nonadgen}) with the results obtained when a specific mechanism is removed from the energy equation. With this analysis, we are able to know where the omitted mechanism has an appreciable effect on the damping. The results of these computations for the fundamental kink modes (Fig.~\ref{fig:mechanismk}) are summarised as follows:

\begin{figure*}[!tb]
\centering
\includegraphics[width=2\columnwidth]{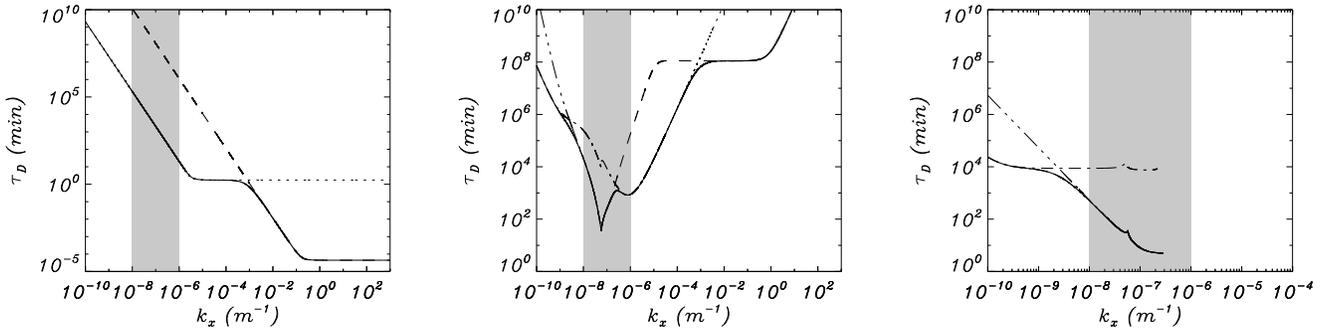}
\caption{Damping time versus the longitudinal wavenumber for the three fundamental kink oscillatory modes: internal slow (left), fast (centre) and external slow (right). Different linestyles represent the omitted mechanism: all mechanisms considered (solid line), prominence conduction eliminated (dotted line), prominence radiation eliminated (dashed line), coronal conduction eliminated (dot-dashed line) and coronal radiation eliminated (three dot-dashed line). Prominence~(1) radiation conditions have been taken for the prominence plasma and the heating scenario is given by $a=b=0$. \label{fig:mechanismk}}
\end{figure*}

\begin{itemize}
\item The fundamental internal slow kink mode is not affected by the mechanisms related with the corona. This is a consequence of the nature of this mode, which propagates strictly along the prominence without disturbing the corona (see Fig. 4, top row, of Paper I). For this reason, in the adiabatic case it is also independent of the coronal conditions. On the other hand, the prominence-related mechanisms show different effects in two different ranges of $k_x$. For \mbox{$k_x \lesssim 10^{-3}\, \mathrm{m}^{-1}$} prominence radiation dominates, while for \mbox{$k_x \gtrsim 10^{-3}\, \mathrm{m}^{-1}$} prominence conduction is the dominant mechanism. Beginning from small values of the wavenumber, prominence radiation becomes more efficient as $k_x$ grows and the damping time falls following a power law until $k_x \approx 10^{-5}\, \mathrm{m}^{-1}$, where $\tau_{\rm D}$ saturates in a plateau between $k_x \approx 10^{-5}$ and $k_x \approx 10^{-3}\, \mathrm{m}^{-1}$. Then, prominence conduction becomes the dominant mechanism and the damping time falls again until $k_x \approx 10^{-1}\, \mathrm{m}^{-1}$ where a new plateau begins. This last part of the curve corresponds to the isothermal or superconductive regime, in which the amplitude of the temperature perturbation drops dramatically (Carbonell et al. \cite{spatial}). Prominence radiation is responsible for the attenuation of the slow mode in the observed wavelength range. An approximate dispersion relation for the internal slow modes is included in App.~\ref{ap:approxslow}.

\item The fundamental fast kink mode is affected by the four mechanisms. For $k_x \lesssim 3 \times 10^{-9}\, \mathrm{m}^{-1}$ coronal radiation dominates but for $3 \times 10^{-9}\, \mathrm{m}^{-1} \lesssim k_x \lesssim 5 \times 10^{-7}\, \mathrm{m}^{-1}$ the effect of coronal conduction grows and becomes the main damping mechanism. Then, for $k_x \gtrsim 5 \times 10^{-7}\, \mathrm{m}^{-1}$ the corona loses dramatically its influence and prominence mechanisms become responsible for the attenuation of this mode. First, prominence radiation is dominant in the range $5 \times 10^{-7}\, \mathrm{m}^{-1} \lesssim k_x \lesssim 10^{-3}\, \mathrm{m}^{-1}$, then prominence conduction governs the wave damping for $k_x \gtrsim 10^{-3}\, \mathrm{m}^{-1}$ and finally the isothermal regime begins for $k_x \approx 10^{0}\, \mathrm{m}^{-1}$. The minimum of $\tau_{\rm D}$ occurs into the coronal conduction regime, for the value of $k_x$ which corresponds to the coupling with the external slow mode. The transition between the coronal conduction regime and the prominence radiation regime occurs in the observed wavelength range.  The 
reason for the sensitivity of the fast mode damping time on prominence and 
coronal conditions is that this wave has a considerable amplitude both inside 
the prominence and in the corona, the later becoming more important for long 
wavelengths (see the second and third rows of Fig. 4 of Paper I).

\item The behaviour of the damping time of the fundamental external slow kink mode is entirely dominated by coronal mechanisms whereas the prominence mechanisms do not affect it at all. This 
behaviour is a result of the negligible amplitude of this wave in the 
prominence (see the fourth and fifth rows of Fig. 4 of Paper I).  For $k_x \lesssim 3 \times 10^{-9}\, \mathrm{m}^{-1}$ coronal radiation dominates, but for shorter wavelengths coronal conduction becomes more relevant and is responsible for the damping in the observed wavelength range until the frequency cut-off is reached. At the cut-off, $\tau_{\rm D}$ has a value of the order of the period. 
\end{itemize}

\begin{figure}[!htb]
\centering
\includegraphics[width=0.75\columnwidth]{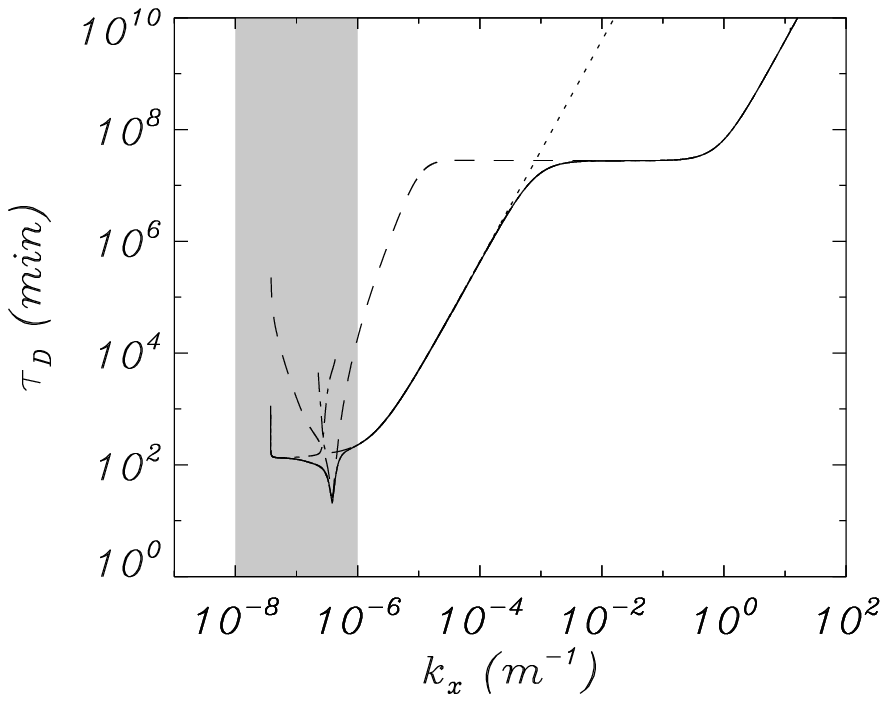}
\caption{Same as Fig.~\ref{fig:mechanismk} for the fundamental fast sausage mode. \label{fig:mechanisms}}
\end{figure}

Regarding the fundamental sausage modes, the behaviour of the internal slow sausage mode is exactly that of the slow kink mode, so no additional comments are needed. The fundamental fast sausage mode (Fig.~\ref{fig:mechanisms}) presents the same scheme as the fundamental fast kink mode for $k_x \gtrsim 10^{-8}$~m$^{-1}$. The main difference between the fast kink and sausage modes happens in the observed wavelength range, where the effect of coronal conduction on the sausage mode is less relevant. If coronal conduction is omitted, the fundamental fast sausage mode is not able to traverse the forbidden region in the dispersion diagram and then shows frequency cut-offs as in the adiabatic case. This means that coronal conduction causes the fast mode to cross the forbidden region in the dispersion diagram in the non-adiabatic case.

\begin{figure*}[!htb]
\centering
\includegraphics[width=2\columnwidth]{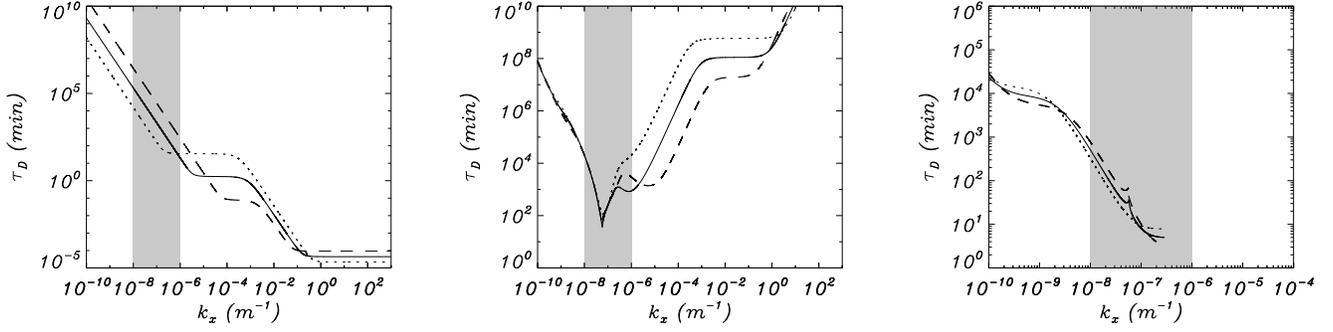}
\caption{Damping time versus the longitudinal wavenumber for the fundamental internal slow kink mode (left), the fundamental fast kink mode (centre) and the fundamental external slow kink mode (right). The different linestyles represent different values of the prominence temperature: $T_{\rm p} = 8000$~K (solid line), $T_{\rm p} = 5000$~K (dotted line) and $T_{\rm p} = 13000$~K (dashed line). The heating scenario is given by $a=b=0$ and the optical thickness for the prominence plasma is Prominence~(1). \label{fig:temp}}
\end{figure*}

\begin{figure*}[!htb]
\centering
\includegraphics[width=2\columnwidth]{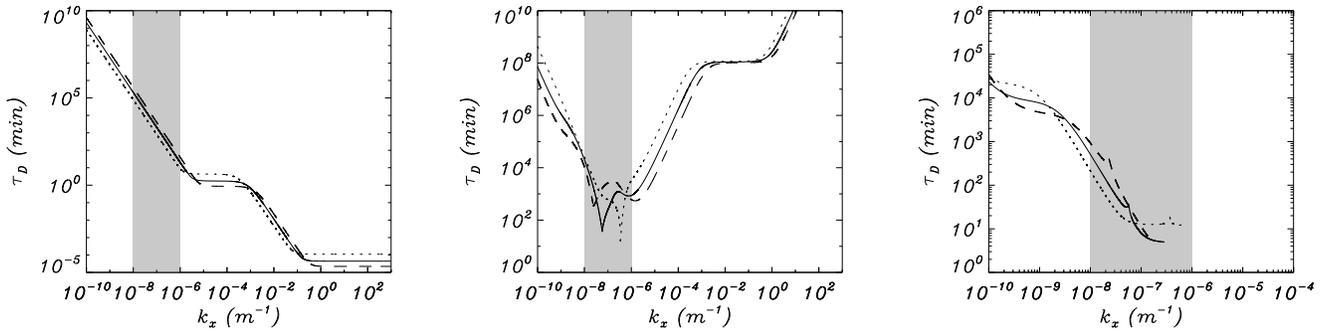}
\caption{Same as Fig.~\ref{fig:temp} with $\rho_{\rm p} = 5 \times 10^{-11}$~kg~m$^{-3}$ (solid line), $\rho_{\rm p} = 2 \times 10^{-11}$~kg~m$^{-3}$ (dotted line) and $\rho_{\rm p} =  10^{-10}$~kg~m$^{-3}$ (dashed line). \label{fig:dens}}
\end{figure*}

\begin{figure*}[!htb]
\centering
\includegraphics[width=2\columnwidth]{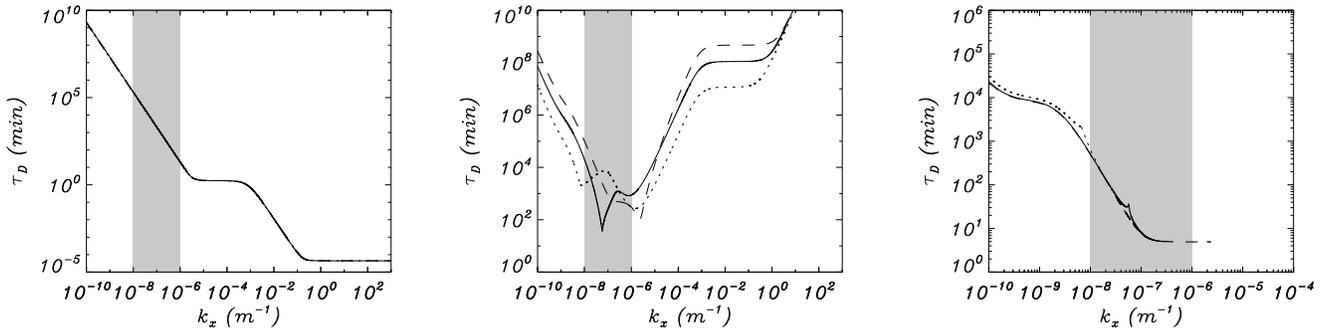}
\caption{Same as Fig.~\ref{fig:temp} with $B_0 = 5$~G (solid line), $B_0 =  2$~G (dotted line) and $B_0 =  10$~G (dashed line). \label{fig:camp}}
\end{figure*}

Approximate values of $k_x$ for which the transitions between regimes take place can be computed by following a process similar to that in Carbonell et al. (2006). The thermal ratio, $d$, and the radiation ratio, $r$, quantify the importance of thermal conduction and radiation, respectively (De Moortel \& Hood 2004),
\begin{eqnarray}
 d & = & \frac{( \gamma -1 ) \kappa_\parallel T_0 \rho_0}{\gamma^2 p_0^2 \tau_{\rm s}} = \frac{1}{\gamma} \frac{\tau_{\rm s}}{\tau_{\rm cond}}, \\
 r & = & \frac{(\gamma -1) \tau_{\rm s} \rho_0^2 \chi^* T_0^\alpha}{\gamma p_0} = \frac{\tau_{\rm s}}{\tau_{\rm rad}},
\end{eqnarray}
where $\tau_{\rm s}$ is the sound travel time and $\tau_{\rm cond}$ and $\tau_{\rm rad}$ are characteristic conductive and radiative time scales. Taking $\tau_{\rm s} = 2 \pi/k^*\cs$, the value of $k^*$ for which the condition $d=r$ is satisfied is
\begin{equation}
 k^* = 2 \pi \rho_0 \sqrt{\frac{ \chi^* T_0^{\alpha-1}}{\kappa_\parallel}}. \label{eq:kp}
\end{equation}
Now, we use prominence values to compute  $k^*$ for the prominence radiation--prominence conduction transition ($k_{\rm p}^*$), and coronal values for the coronal radiation--coronal conduction transition ($k_{\rm c}^*$). This gives the values $k^*_{\rm p} \approx 1.7 \times 10^{-3}$~m$^{-1}$, and $k^*_{\rm c} \approx 2.2 \times 10^{-8}$~m$^{-1}$. For the transition of the fast kink mode between the coronal conduction and the prominence radiation regimes, the boundary wavenumber $k_{\rm p \leftrightarrow c}^*$ can be roughly calculated by imposing $d_{\rm c} = r_{\rm p}$, that gives
\begin{equation}
k_{\rm p \leftrightarrow c}^* = 2 \pi  \rho_{\rm p} \sqrt{\frac{\csco \chi_{\rm p}^* T_{\rm p}^{\alpha_{\rm p}}}{\csp \kappa_{\parallel \rm c} T_{\rm c}}}, 
\end{equation}
and whose numerical value is $k_{\rm p \leftrightarrow c}^* \approx 1.4 \times 10^{-6}$~m$^{-1}$. All these wavenumbers for the transitions between different regimes are independent of the wave type, be it fast or slow, internal or external (this agrees with Figs. \ref{fig:mechanismk} and \ref{fig:mechanisms}). On the other hand, the beginning of the isothermal regime can be estimated by following Porter et al. (1994). Considering $\csp^2 / \vap^2 \ll 1$ and the approximations $\omega_{\rm R} \approx k_x \csp$ for the slow wave and $\omega_{\rm R} \approx k_x \vap$ for the fast wave, the critical wavenumber is
\begin{equation}
 k_{\rm crit-slow} = \frac{2 \rho_{\rm p} k_{\rm B} \csp}{\kappa_{\parallel \rm p} m_{\rm p} \cos \theta},
\end{equation}
for the internal slow mode, and 
\begin{equation}
 k_{\rm crit-fast} = \frac{2 \rho_{\rm p} k_{\rm B} \vap}{\kappa_{\parallel \rm p} m_{\rm p} \cos^2 \theta},
\end{equation}
for the fast mode, where $m_{\rm p}$ is the proton mass, $k_{\rm B}$ is the Boltzmann constant and $\theta$ is the angle between $\vec B$ and $\vec k$. Taking $\cos \theta = 1$ for simplicity, the approximate critical values are $k_{\rm crit-slow} \approx 1.7 \times 10^{-1}$~m$^{-1}$ and $k_{\rm crit-fast} \approx 9.1 \times 10^{-1}$~m$^{-1}$. We note that all these approximate values describe correctly the transitions between the diverse regimes shown in Figs.~\ref{fig:mechanismk} and \ref{fig:mechanisms}, but their numerical values overestimate by almost an order of magnitude the actual critical wavenumbers.


\subsection{Exploring the parameter space}

\begin{figure*}[!tb]
\centering
\includegraphics[width=2\columnwidth]{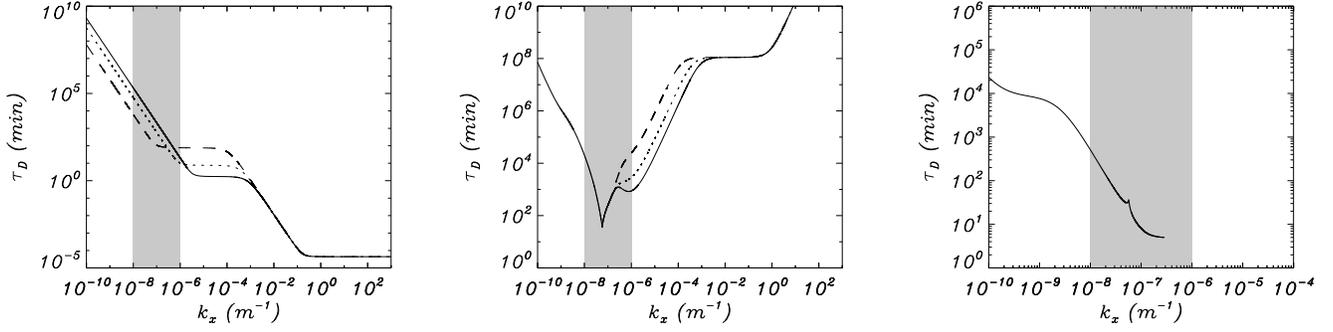}
\caption{Same as Fig.~\ref{fig:temp} with the prominence optical thickness given by Prominence~(1) (solid line), Prominence~(2) (dotted line) and Prominence~(3) (dashed line) conditions. \label{fig:thickness}}
\end{figure*}

\subsubsection{Dependence on the equilibrium physical conditions}
\label{subsec:param}

In this section, we compute the solutions for different values of the equilibrium physical conditions. We only present the results for the fundamental kink modes since they are equivalent to those of sausage modes. Figures~\ref{fig:temp}, \ref{fig:dens} and \ref{fig:camp} display the damping time as function of $k_x$ for some selected values of the prominence temperature, the prominence density and the magnetic field, respectively.

For the internal slow mode, a decrease of the prominence temperature or the prominence density raises the position of the radiative plateau and increases its length. The opposite behaviour is seen when the density or the temperature are increased. However, the value of the magnetic field does not influence the attenuation of this mode, such as expected for a slow wave.

Increasing the value of the prominence temperature causes a vertical displacement of $\tau_{\rm D}$ of the fast mode in those regions in which prominence mechanisms dominate. The value of the prominence density has a smaller effect and its main influence is in changing the coupling point with the external slow mode, which moves to higher $k_x$ for greater values of the density. The magnetic field strength has a more complex effect on $\tau_{\rm D}$ and also modifies the coupling point. 

Finally, the external slow mode is only slightly affected by a modification of the prominence physical parameters since it is mainly dominated by coronal conditions, and the influence of the magnetic field is very small due to the slow-like magnetoacoustic character of this solution.


\subsubsection{Dependence on the prominence optical thickness}
\label{subsec:thickness}

\begin{figure}[!htb]
\centering
\includegraphics[width=0.75\columnwidth]{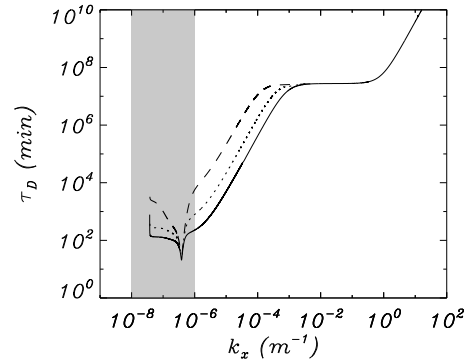}
\caption{Same as Fig.~\ref{fig:thickness} for the fundamental fast sausage mode.  \label{fig:thicknesspekeris}}
\end{figure}

The optically thin radiation assumption is a reasonable approximation in a plasma with coronal conditions but prominence plasmas often are optically thick. In this section we compare the results obtained considering different optical thicknesses for the prominence plasma (see Fig.~\ref{fig:thickness} for the fundamental kink modes).  The results corresponding to the slow sausage mode have not been plotted since they are equivalent to those obtained for slow kink mode; those for the fundamental fast sausage mode, however, are displayed in Fig.~\ref{fig:thicknesspekeris}. 

The variation of the prominence optical thickness modifies the prominence conduction--prominence radiation critical wavenumber, $k^*_{\rm p}$ (see analytical approximation of Eq.~[\ref{eq:kp}]). For the internal slow mode, an increase in the optical thickness raises the position of the radiative plateau and shifts it to smaller wavenumbers. This fact causes an {\em a priori} surprising result in the observed wavelength range, since $\tau_{\rm D}$ has a smaller value for optically thick radiation, Prominence~(3), than for optically thin radiation, Prominence~(1). Regarding fast modes, the damping time increases when the optical thickness is increased, but only in the region in which prominence radiation dominates. The value of $\tau_{\rm D}$ inside the observed wavelength range is partially affected and raises an order of magnitude for Prominence~(3) conditions in comparison with the results for Prominence~(1) conditions. Finally, the damping time of the external slow mode is not affected by the prominence optical thickness since it is entirely dominated by the corona, as it has been noticed in Sect.~\ref{subsec:mechanism}.


\begin{figure*}[!tb]
\centering
\includegraphics[width=2\columnwidth]{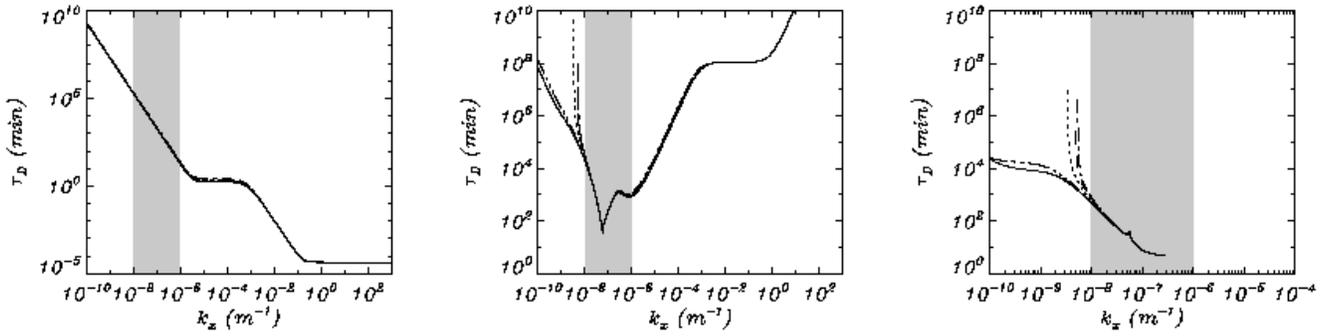}
\caption{Same as Fig.~\ref{fig:temp} with the heating scenario given by $a=b=0$ (solid line); $a=1$, $b=0$ (dotted line); $a=b=1$ (dashed line); $a=b=7/6$ (dot-dashed line); $a=1/2$, $b=-1/2$ (three dot-dashed line). \label{fig:heating}}
\end{figure*}

\subsubsection{Dependence on the heating scenario}
\label{subsec:heating}

 Now, we compute the damping time for the five possible heating scenarios. For simplicity, we only consider the fundamental kink modes (Fig.~\ref{fig:heating}). Carbonell et al. (\cite{carbonell}) showed that in a plasma with prominence conditions the different heating scenarios have no significant influence on the damping time. Nevertheless, in coronal conditions wave instabilities can appear depending on the heating mechanism. In our results, we see that the heating scenario affects the value of $\tau_{\rm D}$ only in the ranges of $k_x$ in which radiation is the dominant damping mechanism. The heating scenario has a negligible effect when prominence radiation dominates, since $\tau_{\rm D}$ is only slightly modified. On the contrary, wave instabilities appear in those regions in which coronal radiation dominates. Thermal destabilisation occurs when the imaginary part of the frequency becomes negative, so oscillations are not attenuated but amplified in time. Instabilities only occur in the fundamental fast kink and the external slow modes for very small values of $k_x$, outside the observed wavelength range.

\subsection{Comparison with the solution for an isolated slab}
\label{subsec:comparision}

In order to assess the effects arising from the presence of two different media in the equilibrium, a comparison between the previous results and those corresponding to a single medium is suitable. So, we consider a simpler equilibrium made of an isolated prominence slab with the magnetic field parallel to its axis. The external medium is not taken into account. Magnetoacoustic non-adiabatic perturbations are governed by Eq.~(\ref{21}), and rigid boundary conditions for $v_z$ are imposed at the edges of the prominence slab,
\begin{equation}
v_z ( -z_{\rm p} ) = v_z ( z_{\rm p} ) = 0.    \label{boundaryslab}
\end{equation}
Then, the solution is of the form
\begin{equation}
  v_z(z) = C_1 \cos \left( \kzp z \right) + C_2 \sin \left( \kzp z \right),  \label{solslab}
\end{equation}
and after imposing boundary conditions (Eq.~[\ref{boundaryslab}]), we deduce the dispersion relation for the magnetoacoustic slow and fast non-adiabatic waves,
\begin{equation}
\kzp z_{\rm p} = \left( n + \frac{1}{2} \right) \pi, \qquad  ( n = 0,1,2,\dots ), \label{slabkink}
\end{equation}
for the kink modes, and
\begin{equation}
\kzp z_{\rm p} = n \pi, \qquad  ( n = 1,2,3,\dots ), \label{slabsausage}
\end{equation}
for the sausage modes. Inserting expressions (\ref{22}) and (\ref{23}) for $\kzp$ and $\lambdap$ respectively, one can rewrite the dispersion relations (\ref{slabkink}) and (\ref{slabsausage}) as polynomial equations in $\omega$. See App.~\ref{ap:slab} for the details.

\begin{figure}[!tb]
\centering
\includegraphics[width=\columnwidth]{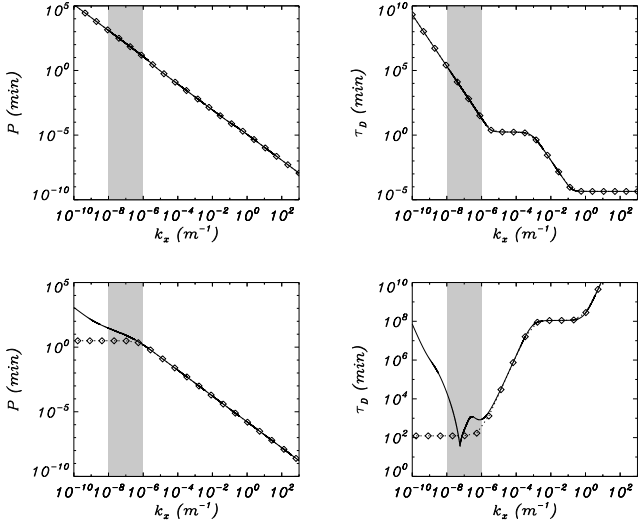}
\caption{Comparison between the solutions for a prominence plus corona system and for an isolated slab with prominence conditions. The upper panels correspond to the fundamental internal slow kink mode and the lower panels to the fundamental fast kink mode. The solid lines are the solutions for a prominence plus corona equilibrium whereas the dotted lines with diamonds represent the solutions for an isolated slab. Prominence~(1) parameters and $a=b=0$ have been used in the computations. \label{fig:slab}}
\end{figure}

Next, considering only the fundamental kink modes for simplicity, we compute the period and the damping time and compare with those obtained when the surrounding corona is taken into account (Fig.~\ref{fig:slab}). We see that there is a perfect agreement between both results in the case of the internal slow mode, whereas the solutions for the fast mode only coincide for intermediate and large wavenumbers, and show an absolutely different behaviour in the observed wavelength range and for smaller wavenumbers. Additionally, one must bear in mind that the external slow mode exists because of the presence of the coronal medium, hence it is not supported by an isolated slab.  

In Paper~I we proved that the internal slow mode is essentially confined within the prominence slab and that the effect of the corona on its oscillatory period can be neglected. Now, we see that the corona has no influence on the damping time either. On the other hand, the confinement of the fast mode is poor for small wavenumbers, the isolated slab approximation not being valid. As it has been noted in Section~\ref{subsec:mechanism}, the corona has an essential effect on the attenuation of the fast mode in the observed wavelength range.

\subsection{Application to a prominence fibril}
\label{subsec:fibril}

Since magnetic field lines are orientated along fibrils, our model can also be applied to study the oscillatory modes supported by a single prominence fibril. In order to perform this investigation, we reduce the slab half-width, $z_{\rm p}$, to a value according to the typical observed size of filament threads, which is between 0.2 to 0.6 arcsec (Lin et al. 2005). Since these values are close to the resolution limit of present-day telescopes, it is likely that thinner threads could exist. So, assuming now $z_{\rm p} = 30$~km, we compute $P$, $\tau_{\rm D}$ and $\tau_{\rm D}/P$ for the fundamental kink modes and compare these results with those obtained for $z_{\rm p} = 3000$~km.

\begin{figure*}[!tb]
\centering
\includegraphics[width=2\columnwidth]{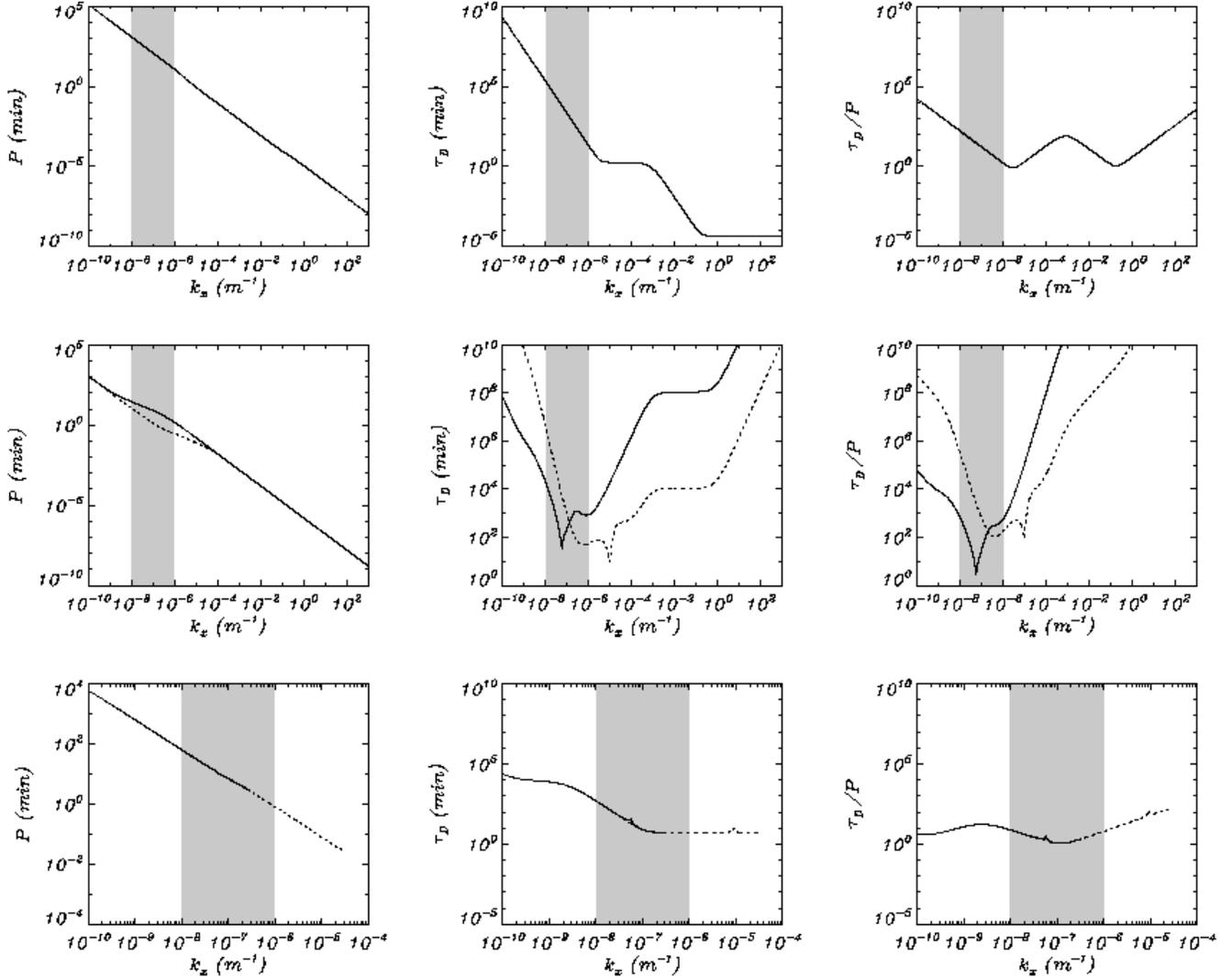}
\caption{ Period (left), damping time (centre) and ratio of the damping time to the period (right) versus $k_x$ for the fundamental kink oscillatory modes: internal slow (top panels), fast (mid panels) and external slow (bottom panels). Solid lines correspond to $z_{\rm p} = 3000$~km whereas dotted lines correspond to $z_{\rm p} = 30$~km. Prominence~(1) radiation conditions have been taken for the prominence plasma and the heating scenario is given by $a=b=0$. \label{fig:fibril}}
\end{figure*}

Such as displayed in Fig.~\ref{fig:fibril}, both internal and external slow modes are not affected by the width of the prominence slab since they are essentially polarised along the $x$-direction and so they are not influenced by the equilibrium structure in the $z$-direction. Nevertheless, the location of the cut-off of the external slow mode and the coupling point with the fast mode are shifted to larger values of $k_x$ when the slab width is reduced. On the other hand, the fast mode, which is responsible for transverse motions, is highly influenced by the value of $z_{\rm p}$. The $\tau_{\rm D}$ curve for the fast mode is displaced to larger values of $k_x$ when smaller $z_{\rm p}$ is considered. This causes that higher values of $\tau_{\rm D}/P$ are obtained in the observed wavelength range. Hence, these results suggest that local prominence oscillations related with transverse fast modes supported by a single fibril could be less affected by non-adiabatic mechanisms than global fast modes supported by the whole or large regions of the prominence. However, according to the results pointed out by D\'iaz et al. (2005) and D\'iaz \& Roberts (2006), large groups of fibrils tend to oscillate together since the separation between individual fibrils is of the order of their thickness.  In a very rough approximation one can consider that a thick prominence slab could represent many near threads which oscillate together and that the larger the slab width, the more threads fit inside it. So, our results show that the slab size (i.e. the number of threads which oscillate together in this rough approximation) has important repercussions on the damping time of collective transverse oscillations, hence the oscillations could be more attenuated when the number of oscillating threads is larger. This affirmation should be verified by investigating the damping in multifibril models.



\section{Conclusions}
\label{sec:conclusions}

In this paper, we have studied the time damping of magnetoacoustic waves in a prominence-corona system considering non-adiabatic terms (thermal conduction, radiation losses and heating) in the energy equation. Small amplitude perturbations have been assumed, so the linearised non-adiabatic MHD equations have been considered and the dispersion relation for the slow and fast magnetoacoustic modes has been found assuming evanescent-like perturbations in the coronal medium. Finally, the damping time of the fundamental oscillatory modes has been computed and the relevance of each non-adiabatic mechanism on the attenuation has been assessed. Next, we summarise the main conclusions of this work:

   \begin{enumerate}
      \item Non-adiabatic effects are an efficient mechanism to obtain small ratios of the damping time to the period in the range of typically observed wavelengths of small-amplitude prominence oscillations.
      
      \item The mechanism responsible for the attenuation of oscillations is different for each magnetoacoustic mode and depends on the wavenumber.
      
      \item The damping of the internal slow mode is dominated by prominence-related mechanisms, prominence radiation being responsible for the attenuation in the observed wavelength range. Such as happens in the adiabatic case (see Paper I) the corona does not affect the slow mode at all, and these results are in perfect agreement with those for an isolated prominence slab.
      
      \item The attenuation of the fast mode in the observed wavelength range is governed by a combined effect of prominence radiation and coronal conduction. The presence of the corona is of paramount importance to explain the behaviour of the damping time for small wavenumbers within the observed range of wavelengths. Non-adiabatic mechanisms in both the prominence 
and the corona are significant because the fast mode achieves large 
amplitudes in both regions.
      
      \item Since the external slow mode is principally supported by the 
corona, its damping time is entirely governed by coronal mechanisms, coronal conduction being the dominant one in the observed wavelength range.

      \item The consideration of different optical thicknesses for the prominence plasma causes an important variation of the damping time of the internal slow and fast modes in the observed wavelength range. Hence a precise knowledge of the radiative processes of prominence plasmas is needed to obtain more realistic theoretical results. 
      
      \item The heating scenario has a negligible effect on the damping time of all solutions in the observed wavelength range. Depending on the scenario considered, thermal instabilities can appear for small values of the wavenumber, in which coronal radiation dominates.
      
	\item The width of the prominence slab does not affect the results for both internal and external slow modes. However, fast modes are less attenuated in the range of observed wavelengths when thinner slabs or filaments threads are considered.

   \end{enumerate}

Taking into account the results in the observed range of wavelengths, one can conclude that radiative effects of the prominence plasma are responsible for the attenuation of the internal slow modes, which can be connected with intermediate- and long-period prominence oscillations, whereas a combined effect of prominence radiation and coronal thermal conduction governs the damping of fast modes, whose periods are compatible with those of short-period oscillations.

\begin{acknowledgements}
     The authors acknowledge the financial support received from the Spanish Minis\-terio de Ciencia y Tecnolog\'ia under grant AYA2006-07637. R.~Soler thanks the Conselleria d'Economia, Hisenda i Innovaci\'o for a fellowship.
\end{acknowledgements}

\appendix

\section{Expressions for the perturbations}

Combining Eqs.~(\ref{15})--(\ref{20}), one can obtain the expressions for the perturbed quantities as functions of $v_z$ and its derivative
\begin{eqnarray}
v_x &=& \frac{- i k_x \Lambda^2}{\omega^2 - k_x^2 \Lambda^2} \frac{{\rm d} v_z}{{\rm d} z}, \\
\rho_1 &=& \frac{i \omega \rho_0}{\omega^2 - k_x^2 \Lambda^2}\frac{{\rm d} v_z}{{\rm d} z}, \\
p_1 &=& \frac{i \omega \rho_0 \Lambda^2}{\omega^2 - k_x^2 \Lambda^2}\frac{{\rm d} v_z}{{\rm d} z}, \\
T_1 &=& \frac{i \omega T_0}{\omega^2 - k_x^2 \Lambda^2} \left( \gamma \frac{\Lambda^2}{\cs^2} - 1 \right)\frac{{\rm d} v_z}{{\rm d} z}, \\
B_{1x} &=& \frac{i B_0}{\omega} \frac{{\rm d} v_z}{{\rm d} z}, \\
B_{1z} &=& \frac{B_0 k_x}{\omega}  v_z.
\end{eqnarray}
Now, we write the expressions for the perturbations to the magnetic pressure, $p_{\rm 1m}$, and the total pressure, $p_{\rm 1T}$,
\begin{eqnarray}
p_{\rm 1m} &=& \frac{B_0}{\mu} B_{1x} = \frac{i \rho_0 \va^2}{\omega} \frac{{\rm d} v_z}{{\rm d} z}, \\
p_{\rm 1T} &=& p_1 + p_{1m} =  \frac{i \rho_0 \left( \omega^2 - k_x^2 \va^2 \right)}{\omega k_z^2} \frac{{\rm d} v_z}{{\rm d} z}.
\end{eqnarray}
In the limit $\Lambda \to \cs$ (i.e. in the absence of conduction, radiation losses and heating), all the expressions reduce to those corresponding to the adiabatic case.

\section{Approximate dispersion relation for the internal slow modes}
\label{ap:approxslow}

\begin{figure}[!htb]
\centering
\includegraphics[width=\columnwidth]{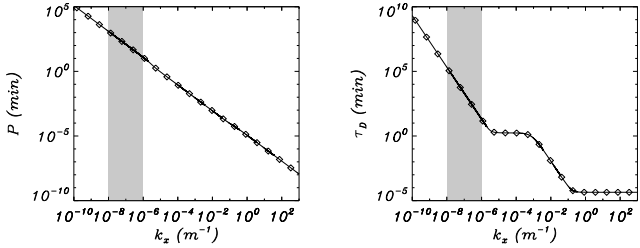}
\caption{Period (left) and damping time (right) versus the longitudinal wavenumber for the fundamental internal slow kink mode. The solid line corresponds to the exact solution and the diamonds correspond to the approximate solution. Prominence~(1) parameters and $a=b=0$ have been used in the computations. \label{fig:slowapprox}}
\end{figure}

Internal slow modes are almost non-dispersive and for adiabatic perturbations a good approximation for the frequency is $\omega \approx \csp k_x$, $\csp$ being the prominence sound speed. In the non-adiabatic case, we can consider the equivalence between $\cs$ and $\Lambda$ to propose $\omega \approx \lambdap k_x$ as an approximate dispersion relation. Taking into account Eq.~(\ref{23}) for $\Lambda$, the approximate dispersion relation for the internal slow modes is a third order polynomial in $\omega$,
\begin{equation}
\omega^3 - i \mathcal{B} \omega^2 - k_x^2 \csp^2 \omega + i \frac{\csp^2}{\gamma} \mathcal{A} k_x^2 = 0,
\end{equation}
with
\begin{eqnarray}
\mathcal{A} &=& \left( \gamma -1 \right) \left( \hat{\kappa}_{\parallel \rm p} k_x^2 + \omega_{T \rm p} - \omega_{\rho \rm p} \right), \label{eqA} \\
 \mathcal{B} &=& \left( \gamma -1 \right) \left( \hat{\kappa}_{\parallel \rm p} k_x^2 + \omega_{T \rm p} \right), \label{eqB} \\
 \hat{\kappa}_{\parallel \rm p} &=& \kappa_{\parallel \rm p} \frac{T_{\rm p}}{p_{\rm p}}. \nonumber
\end{eqnarray}
In Fig.~\ref{fig:slowapprox} a comparison between the exact and approximate solutions is displayed and a perfect agreement is seen.

\section{Dispersion relation for an isolated slab}
\label{ap:slab}

We here deduce a polynomial dispersion relation for the magnetoacoustic normal modes of a slab with a longitudinal magnetic field. Taking Eqs.~(\ref{slabkink}) and (\ref{slabsausage}) as the dispersion relations for the kink and sausage modes, respectively, one can replace $k_z$ and $\Lambda$ with their correspondent expressions (Eqs.~[\ref{22}]--[\ref{23}]), and the following fifth order polynomial equation is found,
\begin{eqnarray}
\omega^5 &-& i \mathcal{B} \omega^4 - \left( \va^2 + \cs^2 \right) \mathcal{K}^2 \omega^3 +  i \left( \va^2 \mathcal{B} +  \mathcal{A} \frac{\cs^2}{\gamma} \right) \mathcal{K}^2 \omega^2  \nonumber \\
&+& \va^2 \cs^2 k_x^2 \mathcal{K}^2 \omega -  \frac{1}{\gamma} i \mathcal{A} \va^2 \cs^2 k_x^2 \mathcal{K}^2 = 0, \label{slabpol}
\end{eqnarray}
with
\begin{displaymath}
\mathcal{K}^2 = k_x^2 + \frac{\left(n + 1/2 \right)^2 \pi^2}{z_{\rm p}^2}, \qquad ( n = 0,1,2,\dots ),
\end{displaymath}
for the kink modes, and
\begin{displaymath}
\mathcal{K}^2 = k_x^2 + \frac{n^2 \pi^2}{z_{\rm p}^2}, \qquad ( n = 1,2,3,\dots ), 
\end{displaymath}
for the sausage modes. Quantities $\mathcal{A}$ and $\mathcal{B}$ are given by Eqs.~(\ref{eqA}) and (\ref{eqB}), respectively.

\end{document}